\documentclass[twocolumn,showpacs,preprintnumbers,amsmath,amssymb,pra,superscriptaddress]{revtex4-1}

\usepackage{multirow}
\usepackage{amsfonts}
\usepackage{amsopn}
\usepackage[english]{babel}
\usepackage[T1]{fontenc}
\usepackage{times}
\usepackage{mathrsfs}
\usepackage{graphicx}
\usepackage{dcolumn}
\usepackage{bm}
\usepackage[colorlinks,bookmarks=true,citecolor=blue,linkcolor=red,urlcolor=blue]{hyperref}
\usepackage[tight,nooneline]{subfigure}
\usepackage{footnote}

\newcommand*{\id}{{\normalfont\hbox{1\kern-0.15em \vrule width .8pt depth-.5pt}}}

\begin{document}
\author{Zhao Liu}
\email{Corresponding author: zhaol@zju.edu.cn}
\affiliation{Zhejiang Institute of Modern Physics, Zhejiang University, Hangzhou 310027, China}
\author{Emil J. Bergholtz}
\affiliation{Department of Physics, Stockholm University, AlbaNova University Center, 106 91 Stockholm, Sweden}
\title{Fractional quantum Hall states with gapped boundaries in an extreme lattice limit}
\date{\today}

\begin{abstract}
We present a detailed microscopic investigation of fractional quantum Hall states with gapped boundaries in a coupled bilayer lattice model featuring holes whose counterpropagating chiral edge states are hybridized and gapped out. We focus on a lattice limit for cold-atom experiments, in which each hole just consists of a single removed site. Although the holes distort the original band structure and lead to ingap remnants of the continuum edge modes, we find that the lowest nearly flat band representing a higher-genus system may naturally form by controlling the local hopping terms that gap out the boundaries. Remarkably, local interactions in this new flat band lead to various Abelian and non-Abelian fractional quantum Hall states with gapped boundaries residing on emergent higher-genus surfaces, which we identify by extracting the nontrivial topological ground-state degeneracies and the fractional statistics of quasiparticles. These results demonstrate the feasibility of realizing novel fractional quantum Hall states with gapped boundaries even in the extreme lattice limit, thus enabling a possible new route towards universal topological quantum computation.
\end{abstract}

\maketitle

\section{Introduction} The standard implementation of topological quantum computation (TQC) \cite{kitaev03,topocompute} relies on non-Abelian anyons and their exotic braiding statistics \cite{moore91,senthil06}. When emerging as quasiparticles in topologically ordered systems \cite{laughlin83,wen91}, these anyons are associated with a topologically protected subspace of degenerate ground states. In this case, the adiabatic braiding of non-Abelian anyons gives rise to unitary transformations in the relevant ground-state subspace and realizes the required fault-tolerant quantum gates.

Apart from braiding non-Abelian anyons, there are also other routes to achieve and even facilitate the realization of TQC. One attractive possibility is to utilize topological states with gapped boundaries \cite{anton,juven,yidun,juven2,iris} (see Refs.~\cite{TCtwists,ady12,you12,qi12,qi13,qi13_2,vaezi13,yy13,brown13,vaezi14,fuchs14,qi15,jiang15,yy16,andrey16,vaezi17,liugenon,guanyu} for another scheme using twist defects which are different from gapped boundaries). These states were predicted to emerge from nonchiral topological orders after multiple disconnected boundaries thereof are gapped out by particle tunneling or pairing between counterpropagating edge modes and can be thought of as residing on an effective higher-genus surface~\cite{barkeshli,chargeproj,ganeshan,cecile}. Gapped boundaries and the emergent higher-genus surface provide access to a richer group of topologically protected unitary transformations than can be realized by braiding anyons alone \cite{chargeproj}. It has been proved that the braiding of gapped boundaries, together with modular transformations in the mapping class group of the effective higher-genus surface \cite{birman} and topological charge measurements, allows universal TQC even though the intrinsic anyons of the underlying phase do not support it \cite{walker,chargeproj,iris2,iris3}. However, the birth and development of the beautiful idea above heavily relied on effective field theory in the continuum~\cite{anton,juven,yidun,juven2,barkeshli,chargeproj}. Given the recent progress on creating and gluing boundaries in optical lattices~\cite{punchole}, a question naturally arises: how well do the descriptions of gapped boundaries in continuum effective field theory apply to lattices? 

In this work, we make a step towards answering this question by using extensive exact diagonalization to search for fractional quantum Hall (FQH) states \cite{otherreview,Emilreview,titusreview} with gapped boundaries in a microscopic lattice model far from the continuum picture. In order to pursue the most pronounced lattice effects, we target minimal boundaries that are created by puncturing single-site holes in a bilayer lattice, combined with high magnetic flux densities piercing the lattice. Remarkably, after gapping out the in-gap vestiges of continuum edge modes by interlayer tunneling, we obtain compelling results, including the ground-state degeneracy and the statistics of quasiparticles, that explicitly demonstrate the existence of lattice FQH states with gapped boundaries residing on effective higher-genus surfaces. Our results thus justify the insights from low-energy field theory in an extreme lattice limit and provide deep insight into the physical realization of FQH states and TQC with gapped boundaries in microscopic lattice models.

\begin{figure*}
\centerline{\includegraphics[width=0.7\linewidth]{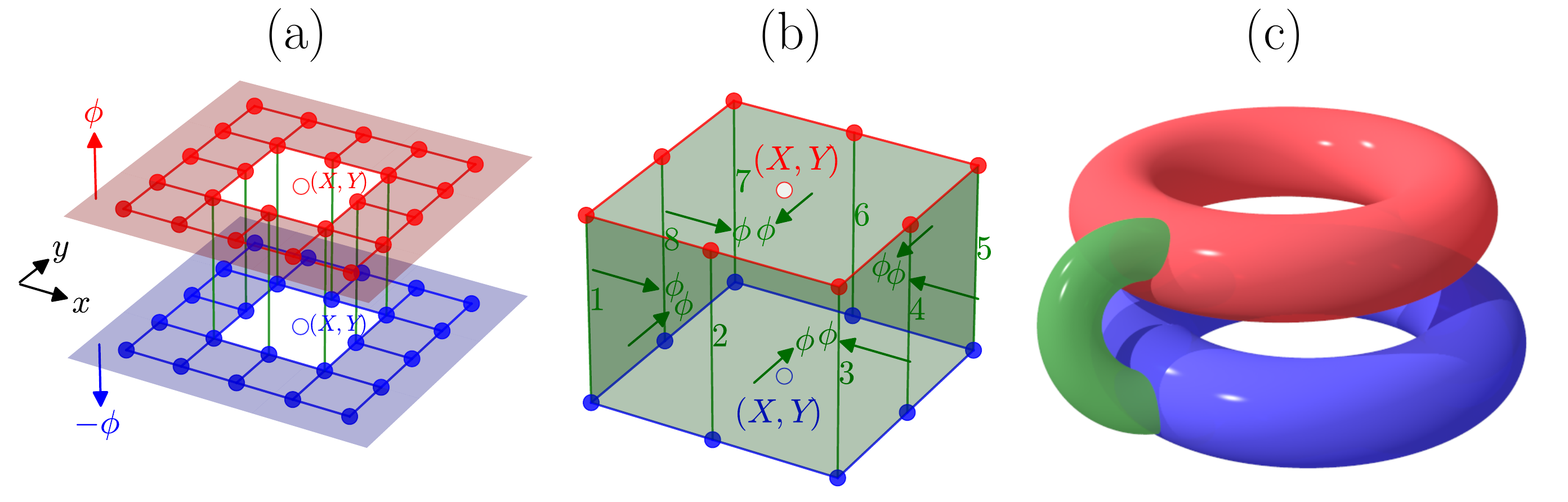}}
\caption{Lattice model with minimal holes and effective higher-genus topology.
(a) Our model includes two square-lattice layers (red and blue), pierced by magnetic flux $\phi$ and $-\phi$ in each plaquette, respectively. We plot only the NN hopping for simplicity. The pair of holes at position $(X,Y)$ (white region) contains a single removed lattice site (open dot) at $(X,Y)$ per layer, such that the two holes are on top of each other. (b) The edges of these two holes are coupled by vertical interlayer tunneling terms $1,\ldots, 8$ (green segment), for which $t^{\perp}_e$ is chosen as $\tilde{t}_2$, $\tilde{t}_1 e^{-2\pi i\phi}$, $\tilde{t}_2 e^{-4\pi i\phi}$, $\tilde{t}_1 e^{2\pi i\phi(2X-1)}$, $\tilde{t}_2 e^{8\pi i\phi X}$, $\tilde{t}_1 e^{2\pi i \phi(4X-1)}$, $\tilde{t}_2 e^{2\pi i\phi(4X-2)}$, and $\tilde{t}_1 e^{2\pi i\phi(2X-1)}$, respectively. Each vertical plaquette (green region) is then pierced by effective flux $\phi$ (green arrow) inwardly. (c) Each pair of holes has the topology of a wormhole, leading to an effective higher-genus surface. 
}
\label{fig:lattice}
\end{figure*}

The remainder of the paper is organized as follows. In Sec.~\ref{sec::model}, we introduce a bilayer lattice model featuring minimal holes and formulate its tight-binding Hamiltonian. This model has small finite-size effects due to the presence of long-range hopping in it. In Sec.~\ref{sec::h0}, we study the effect of holes on the band structure of this long-range hopping model and show that a nearly flat lowest band corresponding to a higher-genus surface can be restored after counterpropagating chiral edge states around holes are gapped out by local interlayer tunneling. In Sec.~\ref{sec::fqh}, we switch on interactions and present solid numerical evidence of bosonic Read-Rezayi states with gapped boundaries. These FQH states are identified by their characteristic ground-state degeneracies and quasiparticle statistics on higher-genus surfaces induced by gapped boundaries. In Sec.~\ref{sec::hofstadter}, in order to consider possible physical realizations of those FQH states with gapped boundaries, we study a bilayer lattice model with only the nearest-neighbor hopping per layer, i.e., Hofstadter bilayers with minimal holes. We find that such a simplified model gives qualitatively the same single-particle and many-body results as the long-range hopping model studied in Secs.~\ref{sec::model}, \ref{sec::h0}, and \ref{sec::fqh}. We summarize our conclusions in Sec.~\ref{sec::conclusion} and show additional data and extensions of our results in the appendixes.

\section{Model} 
\label{sec::model}
We consider a bilayer square lattice in the $xy$ plane, with periodic boundary conditions per layer [Fig.~\ref{fig:lattice}(a)]. In each layer, a lattice site $j$ is labeled by its position $(x_j,y_j)$, where $x_j=0,\ldots, L_x-1$ and $y_j=0,\ldots, L_y-1$.
Each layer is pierced by a uniform magnetic field, such that the number of flux quanta in an elementary plaquette of layer $\sigma$ is $\phi_\sigma$, where $\sigma=\uparrow,\downarrow$. We choose $\phi_\uparrow=-\phi_\downarrow=\phi>0$ to make one layer the time-reversal conjugate of the other. We then punch through both layers to generate $M$ pairs of holes. In each pair, only two lattice sites on top of each other are removed, so the edge of each hole contains four nearest-neighbor (NN) and four next-nearest-neighbor (NNN) sites of the corresponding removed site [Fig.~\ref{fig:lattice}(a)]. Throughout this work, we keep holes in the same layer well separated to avoid overlaps between their edges, and all of our results do not qualitatively depend on specific positions of holes. We couple the edges of each pair of holes with vertical interlayer tunneling [Fig.~\ref{fig:lattice}(b)]. As both layers have the torus geometry, we effectively generate a single surface with genus $g=M+1$ [Fig.~\ref{fig:lattice}(c)].

Based on the scenario described above, we formulate the single-particle Hamiltonian as
\begin{eqnarray}
\label{eq::H0}
H_0=\sum_{j,k\notin \mathcal{R}}\sum_{\sigma=\uparrow,\downarrow}t^\sigma_{jk} a^\dagger_{j\sigma}a_{k\sigma}
+\sum_{m=1}^M\sum_{e\in\mathcal{E}_m}\left(t^{\perp}_e a^\dagger_{e\uparrow}a_{e\downarrow}+{\rm H.c.}\right),\nonumber\\
\end{eqnarray}
where $a^\dagger_{j\sigma}$ ($a_{j\sigma}$) creates (annihilates) a particle in layer $\sigma$ at position $(x_j,y_j)$, $\mathcal{R}$ contains $M$ removed sites in a single layer, and $\mathcal{E}_m$ includes eight edge sites of the $m$th hole in a single layer. For now we choose local intralayer hopping as $t^\sigma_{jk}=t_0(-1)^{x+y+xy} e^{-\frac{\pi}{2}(1-|\phi_\sigma|)(x^2+y^2)} e^{i\pi\phi_\sigma(x_j+x_k)y}$ with $x=x_j-x_k$ and $y=y_j-y_k$~\cite{tjk}, which follows a superexponential decay with the hopping range. As we will show later, while the qualitative physics does not rely on this choice (using the Hofstadter model with only the NN hopping per layer gives similar results), finite-size effects can be suppressed by the tail of $t^\sigma_{jk}$~\cite{ModelFCI}. We set the amplitude of interlayer tunneling ${t}^{\perp}_e$ as $\tilde{t}_1$ ($\tilde{t}_2$) for the NN (NNN) edge sites of each removed site. The phase of ${t}^{\perp}_e$ is fixed to guarantee that each vertical plaquette between a pair of holes is pierced inwardly by effective flux $\phi$ [Fig.~\ref{fig:lattice}(b)], mimicking a magnetic field whose direction is consistent with that in each layer. We focus on $\phi=1/q$ with integer $q>2$, and assume $L_x$ is divisible by $q$ to ensure an integer number of unit cells in the $x$ direction. 

\begin{figure*}
\centerline{\includegraphics[width=\linewidth] {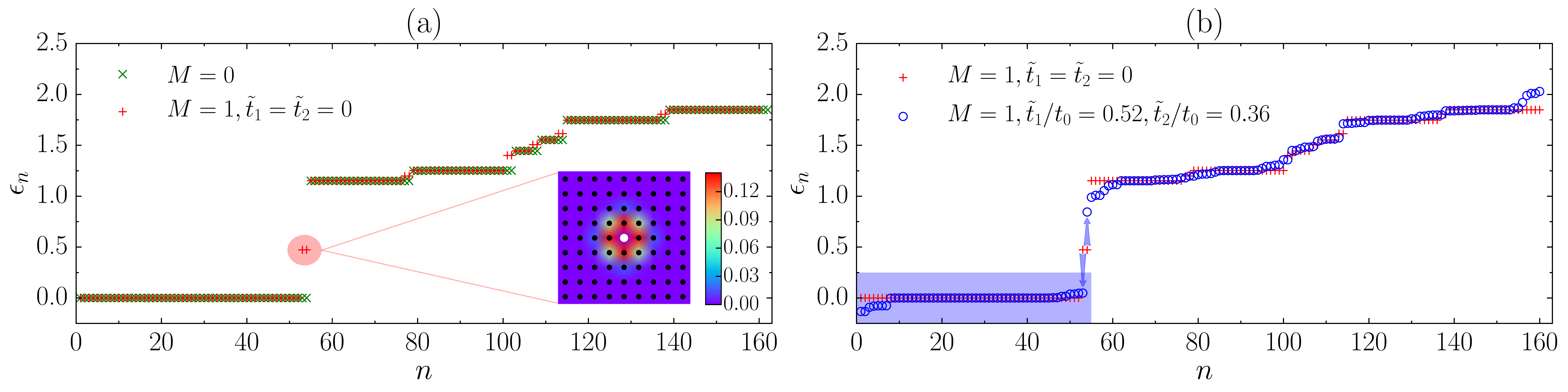}}
\caption{Band structure of Kapit-Mueller bilayers. 
We show the single-particle spectra $\{\epsilon_n\}$ of $H_0$ on an $L_x\times L_y=9\times 9$ lattice with $\phi=1/3$. (a) Without holes (green crosses), there are $2\phi L_xL_y=54$ zero-energy levels. After puncturing $M=1$ pair of holes without interlayer tunneling (red pluses), $2\phi L_xL_y-2M=52$ levels stay at zero energy, while $2M=2$ levels move into the lowest band gap. These two ingap states (red shading) are localized in different layers, where they have identical lattice-site weights concentrating around holes (inset, with the white dot denoting a removed lattice site). (b) With nonzero interlayer tunneling (blue circles), one of the two ingap levels goes up, but the other drops down, as indicated by blue arrows. For suitable tunneling strength, we get a new lowest flat band, consisting of $2\phi L_xL_y-M=53$ states (blue shading).}
\label{fig::SP_hole}
\end{figure*}

\section{Single-particle spectrum and higher-genus flat band} 
\label{sec::h0}
We diagonalize $H_0$ to analyze the effect of holes on the band structure. In the absence of holes, as $H_0$ corresponds to two decoupled Kapit-Mueller models~\cite{kapit} with opposite chiralities, its lowest $2\phi L_xL_y$ eigenstates are exactly degenerate at zero energy \cite{kapit}. However, in the presence of $M$ pairs of holes without interlayer tunneling, while the lowest $2\phi L_xL_y-2M$ levels still stay at zero energy, there are $2M$ levels going into the lowest band gap and forming a quasidegenerate cluster. Typical band structures for $M=0$ and $M=1$ are shown in Fig.~\ref{fig::SP_hole}(a). Remarkably, the eigenvectors of those $2M$ ingap states have dominant weights on lattice sites around holes [inset of Fig.~\ref{fig::SP_hole}(a)], indicating that they are the remnants of $M$ pairs of counterpropagating continuum edge modes corresponding to opposite chiralities of two layers.

\begin{table}
\caption{\label{tab::t1t2} Optimal $\tilde{t}_1$ and $\tilde{t}_2$ yielding the largest flatness $f$ of the lowest $2\phi L_xL_y-M$ eigenstates of $H_0$. Here we choose $L_x\times L_y=24\times 24$ for $\phi=1/3$, $1/4$, and $1/6$ and $L_x\times L_y=25\times 25$ for $\phi=1/5$. These values are not sensitive to the number and positions of well-separated holes, and they remain the same on larger lattices.}
\begin{center}
\begin{ruledtabular}
\begin{tabular}{cccc}
$\phi$ & $\tilde{t}_1/t_0$ & $\tilde{t}_2/t_0$ & $f$ \\ \hline
$1/3$ & $0.52$ & $0.36$ & $4.40$ \\ 
$1/4$ & $0.42$ & $0.24$ & $5.81$ \\
$1/5$ & $0.36$ & $0.19$ & $6.94$ \\ 
$1/6$ & $0.33$ & $0.15$ & $7.96$
\end{tabular}
\end{ruledtabular}
\end{center}
\end{table}

When the interlayer tunneling is switched on, the two edge states around each pair of holes are coupled, and the boundaries become gapped. Indeed, compared to the zero-tunneling case, $M$ in-gap levels are pushed to higher energies, but the other $M$ ingap states go down, such that a band gap is reopened [Fig.~\ref{fig::SP_hole}(b)]. With suitable tunneling strength, we can obtain a new lowest band of high flatness containing $2\phi L_xL_y-M$ eigenstates of $H_0$. In Table~\ref{tab::t1t2}, we show the optimal interlayer tunneling strengths $\tilde{t}_1$ and $\tilde{t}_2$ that yield the flattest lowest band within our setup. Notably, as the flux density $\phi$ decreases, we can get a flatter lowest band with weaker tunneling. This new lowest flat band corresponds to a higher-genus system on the effective $g=M+1$ surface. Similar single-particle physics can also be obtained for larger holes (see Appendix~\ref{sec::lh}).

\section{FQH states with gapped boundaries} 
\label{sec::fqh}
Having ensured that a new lowest flat band can be recovered after the boundaries are gapped out in Kapit-Mueller bilayers with holes, we now examine whether suitable interactions can stabilize FQH states with gapped boundaries, which should reside on the effective $g=M+1$ surface. Due to the relevance for the cold-atom implementation, we focus on the possibility of bosonic Read-Rezayi (RR) states \cite{rr} at filling $\nu=k/2$ with integer $k>0$. This filling is defined as $\nu=\lim_{N_b\rightarrow\infty}N_b/N_s$, where $N_b$ is the number of bosons and $N_s=2\phi L_xL_y-M$ is the number of single-particle states in the new flat band of $H_0$. $k=1$ and $k=2$ correspond to the Laughlin state~\cite{laughlin83} and the Moore-Read (MR) state \cite{moore91}, respectively. Considering that $N_s=2N_b/k-(1-g)$ holds for the $\nu=k/2$ RR state on a single genus-$g$ surface in the lowest Landau level~\cite{wenshift,liugenon}, where the extra offset $1-g$ is related to the topological ``shift''~\cite{wenshift}, we set $N_b=k(\phi L_xL_y-M)$ in our lattice model on the effective $g=M+1$ surface. We assume that bosons at $\nu=k/2$ interact via  
\begin{eqnarray}
\label{Hint}
H_{\mathrm{int}}=U\sum_{\sigma=\uparrow,\downarrow}\sum_{i\notin \mathcal{R}} :n_{i,\sigma}n_{i,\sigma}\cdots n_{i,\sigma}:,
\end{eqnarray}
where $n_{i,\sigma}$ is the occupation on site $i$ in layer $\sigma$, $U>0$ is the interaction strength, and $:\cdots:$ enforces the normal ordering.

\begin{figure}
\centerline{\includegraphics[width=\linewidth] {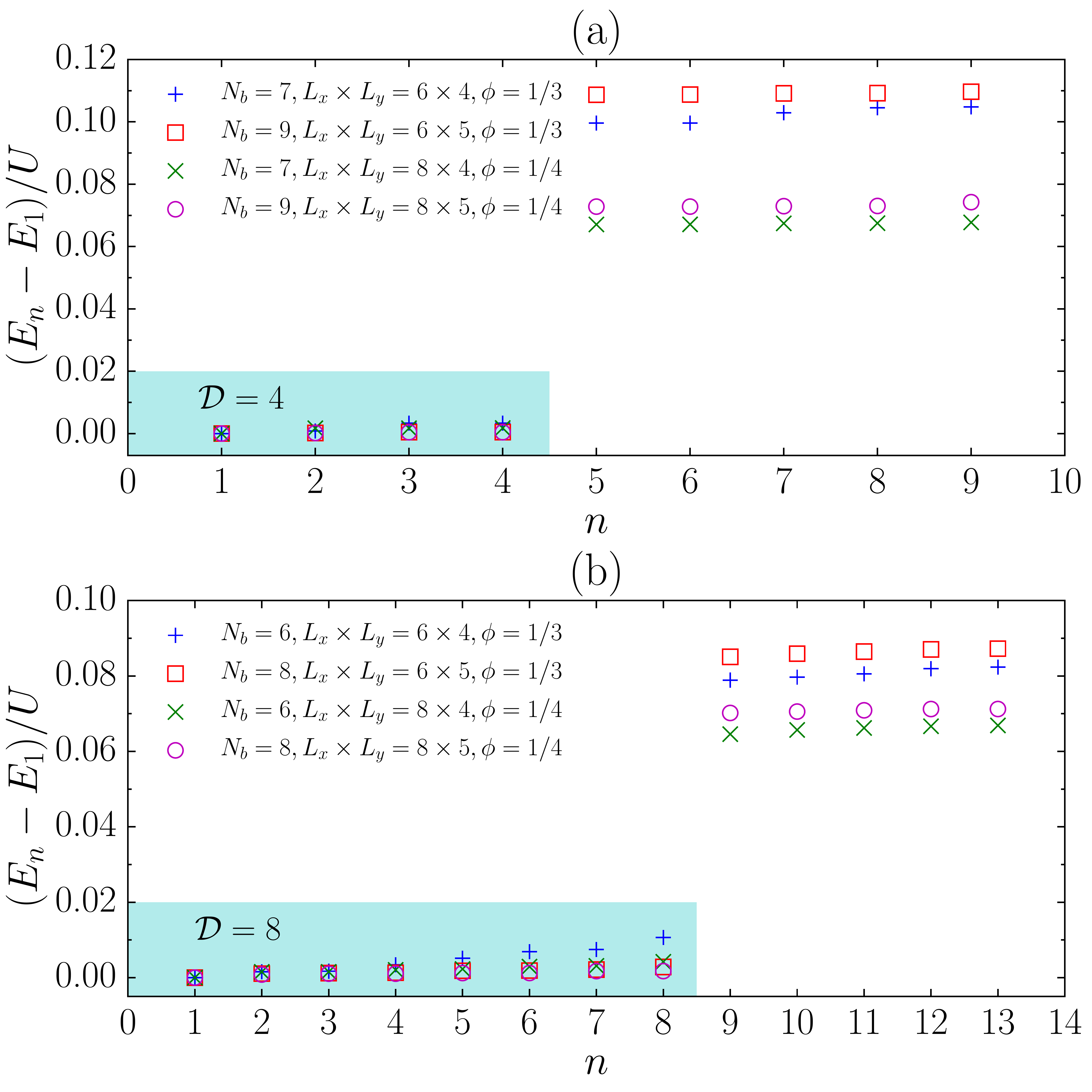}}
\caption{Evidence of Abelian higher-genus FQH states in the band-flattened model of Kapit-Mueller bilayers with holes. We show the spectra $\{E_n\}$ of the interaction projected to the lowest band of $H_0$ at $\nu=1/2$ for various system sizes with (a) $M=1$ and (b) $M=2$ pairs of holes. We use $\tilde{t}_1$ and $\tilde{t}_2$ in Table~\ref{tab::t1t2}. The positions of holes are given in Appendix~\ref{sec::hpos}. The ground-state manifold and the degeneracy $\mathcal{D}$ are highlighted by the cyan shading.}
\label{fig::lau}
\end{figure}

Let us first consider the most realistic case with $k=1$ at $\nu=1/2$. In principle, exactly diagonalizing $H_0+H_{\mathrm{int}}$ in real space provides the low-energy properties of the system. However, because holes break the translation invariance, such calculations are computationally expensive, thus being limited to very small systems. Assuming that $U$ is larger than the band dispersion but smaller than the band gap, we project the interaction $H_{\mathrm{int}}$ to the higher-genus flat lowest band of $H_0$ (see Appendix~\ref{sec::proj}) and neglect the band dispersion for large numerical efficiency. Similar band-flattened models have also been extensively used to study FQH states on translationally invariant lattices \cite{rbprx}. By diagonalizing the projected interaction for $M=1$ and $M=2$, we find compelling evidence that the ground state is the $\nu=1/2$ Laughlin state on the effective $g=M+1$ surface. With one pair of holes, there are four approximately degenerate ground states for various system sizes [Fig.~\ref{fig::lau}(a)], which is consistent with the $\nu=1/2$ Laughlin state on a single $g=2$ surface. A nontrivial enhancement of the topological degeneracy $\mathcal{D}$ from $4$ to $8$ occurs for two pairs of holes [Fig.~\ref{fig::lau}(b)], matching the $\nu=1/2$ Laughlin state on a single $g=3$ surface. For both $M=1$ and $M=2$, the ground states are separated from other excited states by an energy gap which is significantly larger than the ground-state splitting, and the splitting is reduced relative to the gap as the system size is increased.

\begin{figure}
\centerline{\includegraphics[width=\linewidth] {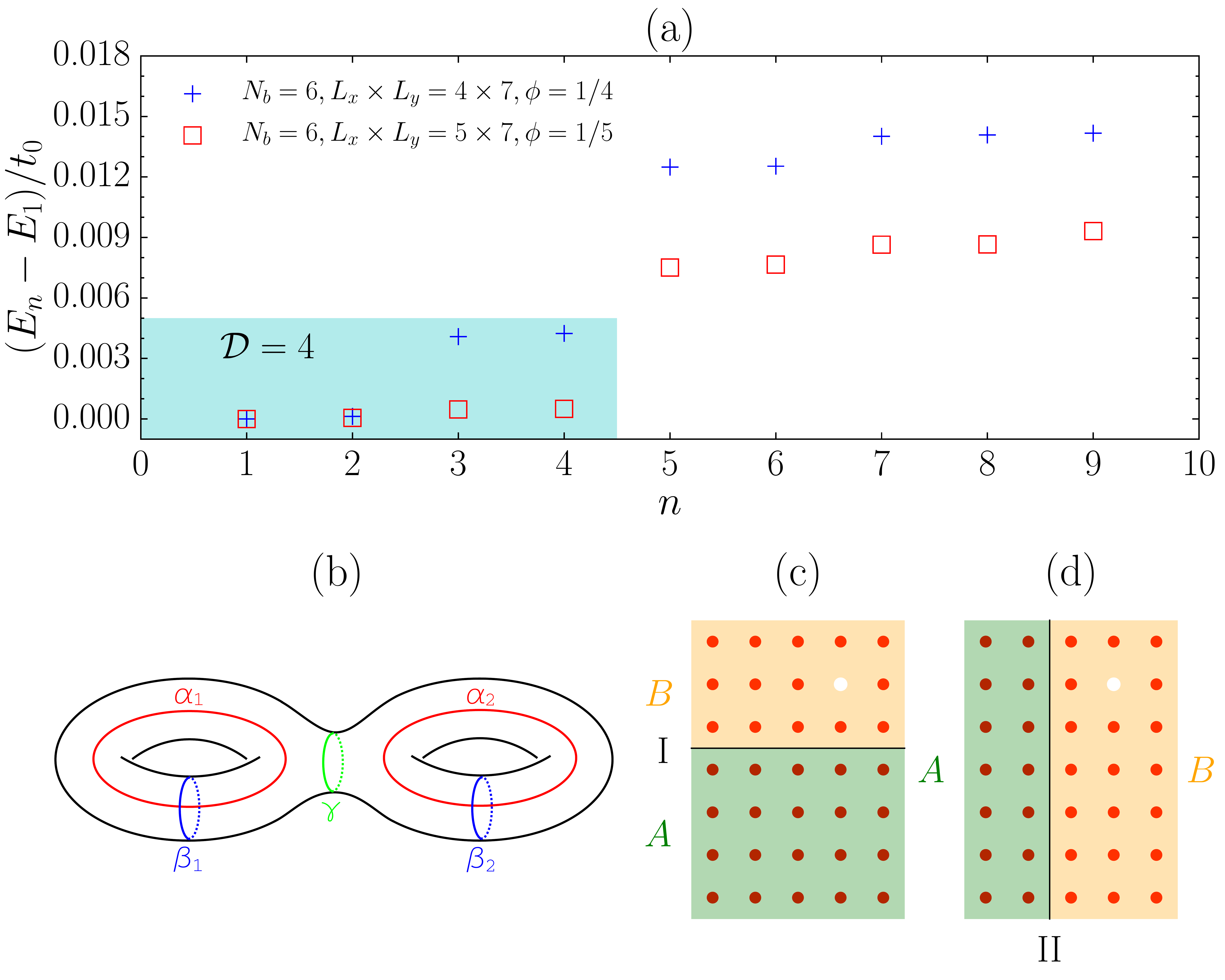}}
\caption{Evidence of Abelian higher-genus FQH states in Kapit-Mueller bilayers with holes, with the band dispersion and mixture taken into account. (a) The spectra $\{E_n\}$ of $H_0+H_{\mathrm{int}}$ for hardcore bosons at $\nu=1/2$ with one pair of holes at position $(X,Y)$. Here we use $\tilde{t}_1/t_0=0.23,\tilde{t}_2/t_0=0.41,(X,Y)=(2,5)$ for $L_x\times L_y=4\times 7,\phi=1/4$ and $\tilde{t}_1/t_0=0.18,\tilde{t}_2/t_0=0.36,(X,Y)=(3,5)$ for $L_x\times L_y=5\times 7,\phi=1/5$. The ground-state manifold and the degeneracy $\mathcal{D}$ are highlighted by the cyan shading. (b) Some noncontractible circles on a $g=2$ surface. Top view of subsystems $A$ and $B$ generated by (c) cut I and (d) cut II on a lattice with $L_x\times L_y=5\times 7,(X,Y)=(3,5)$. Cuts I and II go through both layers, and correspond to circles $(\alpha_1\alpha_2)$ and $(\beta_1\beta_2)$ in (b), respectively. Note that $A$ and $B$ are bilayer subsystems. The removed sites are represented by white dots.}
\label{fig::lau_hc}
\end{figure} 

\begin{figure*}
\centerline{\includegraphics[width=\linewidth] {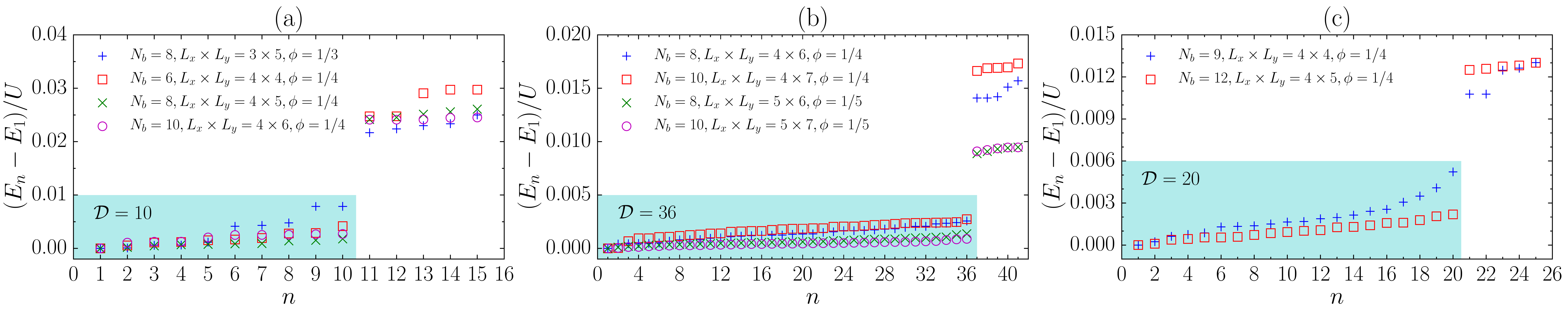}}
\caption{Evidence of non-Abelian higher-genus FQH states in the band-flattened model of Kapit-Mueller bilayers with holes. We show the spectra $\{E_n\}$ of the interaction projected to the lowest band of $H_0$ for various system sizes with (a) $\nu=1,M=1$, (b) $\nu=1,M=2$, and (c) $\nu=3/2,M=1$. We use $\tilde{t}_1$ and $\tilde{t}_2$ in Table~\ref{tab::t1t2}. The positions of holes are given in Appendix~\ref{sec::hpos}. The ground-state manifold and the degeneracy $\mathcal{D}$ are highlighted by the cyan shading.}
\label{fig::nA}
\end{figure*}

\begin{table*}
\caption{\label{tab::mes} The R\'enyi-$2$ entropy $S_2$ of the four MESs with respect to different bipartitions of the $\nu=1/2$ Laughlin state in the
Kapit-Mueller bilayers with one pair of holes. Cuts I and II are shown in Fig.~\ref{fig::lau_hc}. For the cut between two layers, each subsystem contains a whole layer.}
\begin{center}
\begin{ruledtabular}
\begin{tabular}{ccc}
 & $N_b=6,L_x\times L_y=4\times 7,\phi=1/4$ & $N_b=6,L_x\times L_y=5\times 7,\phi=1/5$ \\ \hline
Cut I & $S_2\approx 1.37908,1.36319,1.36319,1.37908$ & $S_2\approx 1.76580,1.71694,1.71694,1.76580$ \\
Cut II & $S_2\approx 2.86280,2.82103,2.91412,2.86280$ & $S_2\approx 3.12519,3.27780,3.27780,3.42259$ \\
Cut between two layers & $S_2\approx 0.357869,0.357887,0.530498,0.536709$ & $S_2\approx 0.322536,0.322536,0.355010,0.355196$
\end{tabular}
\end{ruledtabular}
\end{center}
\end{table*}

To further identify the ground states observed above, we extract their modular $\mathcal{S}$ matrix encoding the quasiparticle statistics \cite{verlinde,wenrigid}. We focus on the effective $g=2$ surface induced by one pair of holes. For an Abelian state on such a surface, the $\mathcal{S}$ matrix relates two specific bases, $\{|acb\rangle_{\alpha_1\gamma\alpha_2}\}$ and $\{|acb\rangle_{\beta_1\gamma\beta_2}\}$, in the ground-state manifold via $|a'cb'\rangle_{\beta_1\gamma\beta_2}=\sum_{a,b}\mathcal{S}_{aa'}\mathcal{S}_{bb'}|acb\rangle_{\alpha_1\gamma\alpha_2}$, where $|acb\rangle_{\alpha_1\gamma\alpha_2}$ ($|acb\rangle_{\beta_1\gamma\beta_2}$) has quasiparticles $a$, $c$, and $b$ threading the nonintersecting, noncontractible circles $\alpha_1$, $\gamma$, and $\alpha_2$ ($\beta_1$, $\gamma$, and $\beta_2$), respectively [Fig.~\ref{fig::lau_hc}(b)], and the quasiparticle $c$ must be $\id$ \cite{chargeproj}. Therefore, the overlap matrix between $|a\id\ b\rangle_{\alpha_1\gamma\alpha_2}$ and $|a\id\ b\rangle_{\beta_1\gamma\beta_2}$ gives $\mathcal{S}\otimes\mathcal{S}$. $|a\id\ b\rangle_{\alpha_1\gamma\alpha_2}$ ($|a\id\ b\rangle_{\beta_1\gamma\beta_2}$) should have the minimal entanglement entropy between two subsystems if we bipartite the whole system along circles $(\alpha_1\alpha_2)$ [$(\beta_1\beta_2)$] \cite{zhangyi,wzhu1,wzhu2}, which in our model correspond to a cut through both layers in the $x$ ($y$) direction. To obtain these minimally entangled states (MESs), we diagonalize the full Hamiltonian $H_0+H_{\mathrm{int}}$ in real space. Because lattice FQH states may persist even for infinitely strong repulsion~\cite{wang1,wang2,stefanos}, we assume hard-core bosons to increase numerical efficiency. Although the tractable systems under this assumption are still quite small compared to the band projection case, we do observe clear fourfold topological degeneracies at $\nu=1/2$~[Fig.~\ref{fig::lau_hc}(a)]. We then measure the entanglement entropy by the R\'enyi-$2$ entropy $S_2=-\ln{\rm Tr}\rho_A^2$, where $\rho_A$ is the reduced density matrix of subsystem $A$, and numerically search for MESs in the fourfold-degenerate ground-state manifold (see Appendix~\ref{sec::mes}). Strikingly, for both cut I in the $x$ direction [Fig.~\ref{fig::lau_hc}(c)] and cut II in the $y$ direction [Fig.~\ref{fig::lau_hc}(d)], we indeed find four almost orthogonal MESs $|\Sigma_{m=1,2,3,4}^{{\rm I/II}}\rangle$ with similar $S_2$ (Table~\ref{tab::mes}), and the overlap matrix $\mathcal{O}_{mn}=\langle\Sigma_m^{{\rm I}}|\Sigma_n^{{\rm II}}\rangle$ is very close to $\mathcal{S}\otimes\mathcal{S}$, where $\mathcal{S}=\frac{1}{\sqrt{2}}
\left(
\begin{array}{cc}
1&1\\
1&-1
\end{array}\right)
$
is the $\mathcal{S}$ matrix of the $\nu=1/2$ Laughlin state~\cite{laughlinS1,laughlinS2,laughlinS3}. Specifically, we get 
\begin{eqnarray}
\label{eq::modularS2}
\mathcal{O}\approx
\left(
\begin{array}{cccc}
0.523&0.525&0.517&0.523\\
0.477&-0.472&0.483&-0.477\\
0.477&0.472&-0.483&-0.477\\
0.523&-0.525&-0.517&0.523
\end{array}\right)
\end{eqnarray}
for $N_b=6,L_x\times L_y=4\times 7,\phi=1/4$ and 
\begin{eqnarray}
\label{eq::modularS}
\mathcal{O}\approx
\left(
\begin{array}{cccc}
0.493&0.494&0.494&0.496\\
0.507&-0.505&0.505&-0.503\\
0.507&0.505&-0.505&-0.503\\
0.493&-0.494&-0.494&0.496
\end{array}\right)
\end{eqnarray}
for $N_b=6,L_x\times L_y=5\times 7,\phi=1/5$.
As the nonzero minimal entanglement entropy between two layers (Table~\ref{tab::mes}) rules out two decoupled copies of the torus $\nu=1/2$ Laughlin state, we conclude that the ground state at $\nu=1/2$ with one pair of holes is the $\nu=1/2$ Laughlin state on the effective $g=2$ surface.

Having confirmed Abelian higher-genus FQH states with gapped boundaries in Kapit-Mueller bilayers with holes, we now consider whether non-Abelian states with gapped boundaries can appear when $k>1$. In order to reach relatively large systems, we again diagonalize the interaction projected to the higher-genus flat lowest band of $H_0$. We indeed observe ground-state degeneracies consistent with the $\nu=1$ and $\nu=3/2$ RR states on the effective $g=M+1$ surface. When $k=2$, $\mathcal{D}=10$ approximately degenerate ground states exist at the bottom of the many-body spectra for $M=1$ [Fig.~\ref{fig::nA}(a)], and adding another pair of holes ($M=2$) increases $\mathcal{D}$ to $36$ [Fig.~\ref{fig::nA}(b)]. These topological degeneracies match the MR state on a single $g=M+1$ surface~\cite{Zkdeg}. When $k=3$, we get $\mathcal{D}=20$ for $M=1$, which is the same as that of the $\nu=3/2$ RR state on a single $g=2$ surface [Fig.~\ref{fig::nA}(c)]~\cite{Zkdeg}. Note that the topological degeneracies observed above become better for larger system sizes. 

\section{Hofstadter bilayers with holes} 
\label{sec::hofstadter}
So far we have considered a single-particle Hamiltonian with local but long-range hopping, i.e., the Kapit-Mueller bilayers. However, we find that similar results can also be obtained if we keep only the NN hopping. We still focus on minimal holes in this section.

\begin{figure*}
\centerline{\includegraphics[width=\linewidth] {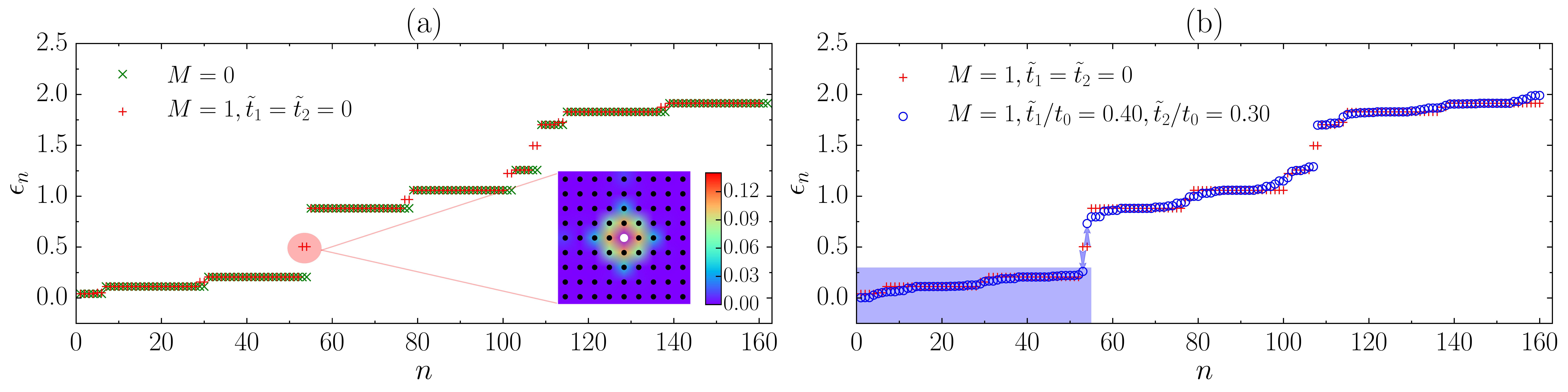}}
\caption{Band structure of Hofstadter bilayers.
We show the spectra $\{\epsilon_n\}$ of $H_0'$ on an $L_x\times L_y=9\times 9$ lattice with $\phi=1/3$. (a) Without holes (green crosses), the lowest band contains $2\phi L_xL_y=54$ levels. After puncturing $M=1$ pair of holes without interlayer tunneling (red pluses), $2\phi L_xL_y-2M=52$ levels stay at their original energies, while $2M=2$ levels move into the lowest band gap. These two ingap states (red shading) are localized in different layers, where they have identical lattice-site weights concentrating around holes (inset, with the white dot denoting a removed lattice site). (b) With nonzero interlayer tunneling (blue circles), one of the two ingap levels goes up, but the other drops down, as indicated by blue arrows. For suitable tunneling strength, we get a new lowest flat band, consisting of $2\phi L_xL_y-M=53$ states (blue shading).}
\label{fig::HS_SP_hole}
\end{figure*}

Under the hopping truncation, we get a new single-particle Hamiltonian
\begin{eqnarray}
\label{eq::H02}
H_0'=\sum_{j,k\notin \mathcal{R}}\sum_{\sigma=\uparrow,\downarrow}t'^\sigma_{jk} a^\dagger_{j\sigma}a_{k\sigma}
+\sum_{m=1}^M\sum_{e\in\mathcal{E}_m}\left(t^{\perp}_e a^\dagger_{e\uparrow}a_{e\downarrow}+{\rm H.c.}\right),\nonumber\\
\end{eqnarray}
which is very similar to $H_0$. The only difference is that now we have $t'^\sigma_{jk}=t_0(-1)^{x+y+xy} e^{-\frac{\pi}{2}(1-|\phi_\sigma|)(x^2+y^2)} e^{i\pi\phi_\sigma(x_j+x_k)y}$ for $x^2+y^2\leq 1$ and $t'^\sigma_{jk}=0$ for $x^2+y^2>1$. In the absence of holes, $H_0'$ corresponds to two decoupled Hofstadter models with opposite chiralities, and the exact flatness of the lowest $2\phi L_x L_y$ eigenstates in the Kapit-Mueller case is lost due to the hopping truncation. However, we still observe an effect of holes on the band structure of $H_0'$ very similar to that for $H_0$: in the presence of $M$ pairs of holes without interlayer tunneling, there are $2M$ levels jumping into the lowest band gap of $H_0'$ and forming a quasidegenerate cluster, while the other $2\phi L_xL_y-2M$ lowest states stay at their original energies [Fig.~\ref{fig::HS_SP_hole}(a)]. Those $2M$ in-gap states are edge states, as their eigenvectors have dominant weights on lattice sites around holes [inset of Fig.~\ref{fig::HS_SP_hole}(a)]. 

Similar to the case with long-range hopping, the interlayer tunneling in $H_0'$ around each pair of holes also pushes $M$ in-gap levels upwards to the higher-energy band, and the other $M$ in-gap levels come down [Fig.~\ref{fig::HS_SP_hole}(b)]. The boundaries are then gapped out, and the band gap is reopened. For suitable tunneling strength, we obtain a new lowest band containing the lowest $2\phi L_xL_y-M$ eigenstates of $H_0'$. In Table~\ref{tab::t1t2nn}, we show the interlayer tunneling strengths $\tilde{t}_1$ and $\tilde{t}_2$, with which this band is the flattest for large enough systems. Compared with the results with long-range hopping (Table~\ref{tab::t1t2}), the maximal band flatness with only the NN hopping is smaller. However, as the flux density $\phi$ decreases, the band also becomes flatter with weaker tunneling. Again, such a new lowest flat band corresponds to a higher-genus system on the effective $g=M+1$ surface.

\begin{figure}
\centerline{\includegraphics[width=\linewidth] {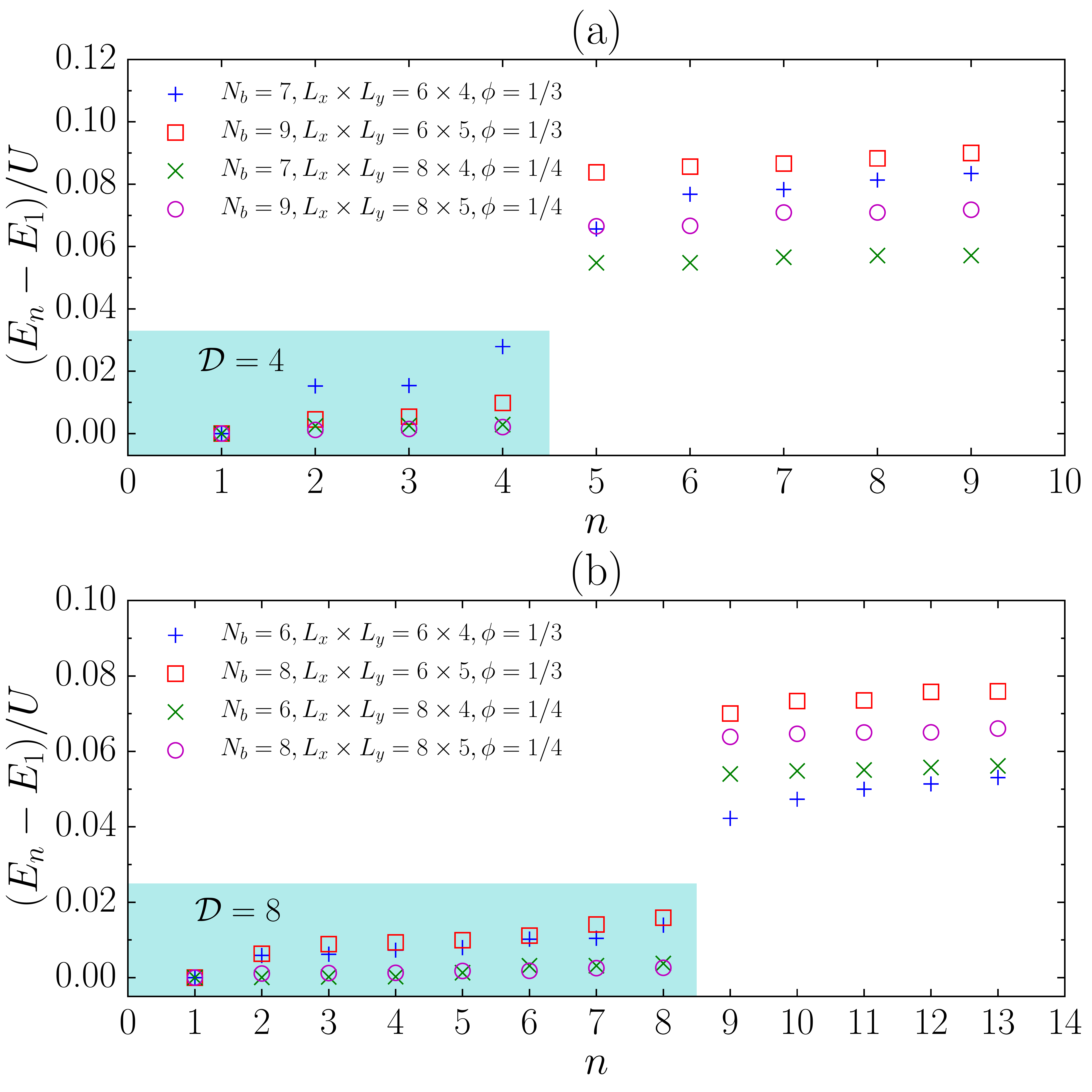}}
\caption{Evidence of Abelian higher-genus FQH states in the band-flattened model of Hofstadter bilayers with holes.
We show the spectra $\{E_n\}$ of the interaction projected to the lowest band of $H_0'$ for various system sizes at $\nu=1/2$ with (a) $M=1$ and (b) $M=2$ pairs of holes. We use $\tilde{t}_1$ and $\tilde{t}_2$ in Table~\ref{tab::t1t2nn}. The positions of holes are given in Appendix~\ref{sec::hpos}. The ground-state manifold and the degeneracy $\mathcal{D}$ are highlighted by the cyan shading. The degeneracies here in general become worse than in Fig.~\ref{fig::lau}.}
\label{fig::HS_lau}
\end{figure}

\begin{figure*}
\centerline{\includegraphics[width=\linewidth] {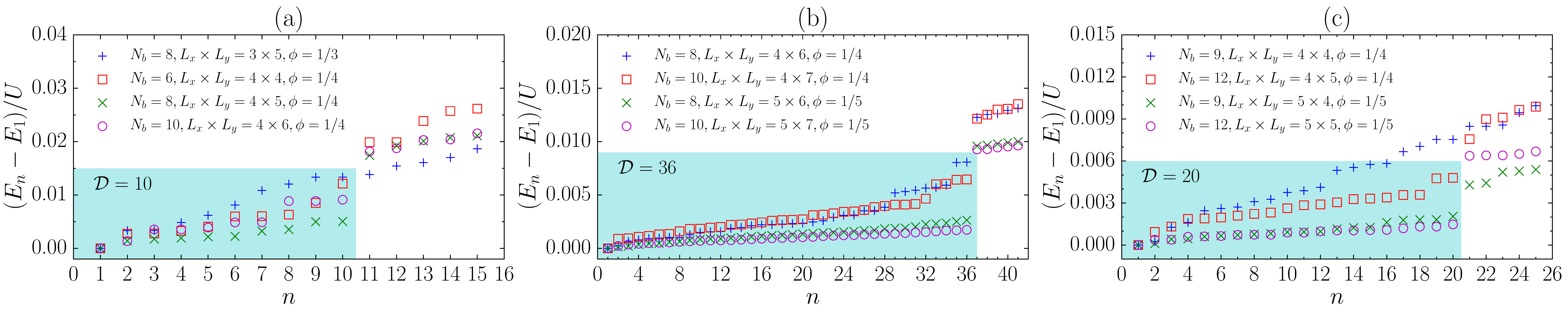}}
\caption{Evidence of non-Abelian higher-genus FQH states in the band-flattened model of Hofstadter bilayers with holes.
We show the spectra $\{E_n\}$ of the interaction projected to the lowest band of $H_0'$ for various system sizes with (a) $\nu=1,M=1$, (b) $\nu=1,M=2$, and (c) $\nu=3/2,M=1$. We use $\tilde{t}_1$ and $\tilde{t}_2$ in Table~\ref{tab::t1t2nn}. The positions of holes are given in Appendix~\ref{sec::hpos}. The ground-state manifold and the degeneracy $\mathcal{D}$ are highlighted by the cyan shading. The degeneracies here in general become worse than in Fig.~\ref{fig::nA}, and even disappear for $N_b=8,L_x\times L_y=3\times 5,\phi=1/3$ in (a) and $N_b=9,L_x\times L_y=4\times 4,\phi=1/4$ in (c).}
\label{fig::HS_nA}
\end{figure*}

We then examine the possibility of the $\nu=k/2$ RR state with gapped boundaries in this new flat band arising from only the NN hopping. We still consider the $(k+1)$-body repulsion. For large numerical efficiency, we project the interaction into the lowest flat band of $H_0'$ and neglect the band dispersion. Although numerical results suggest stronger finite-size effects than in the long-range hopping case, we do find convincing topological ground-state degeneracy consistent with the $\nu=k/2$ RR state on a single higher-genus surface for both $M=1$ and $M=2$ pairs of holes (Figs.~\ref{fig::HS_lau} and \ref{fig::HS_nA}). Our results indicate that the long-range hopping in Eq.~(\ref{eq::H0}) chosen for more rapid finite-size convergence is not essential for stabilizing FQH states with gapped boundaries. 

\begin{table}
\caption{\label{tab::t1t2nn} Optimal $\tilde{t}_1$ and $\tilde{t}_2$ yielding the largest flatness $f$ of the lowest $2\phi L_xL_y-M$ eigenstates of $H_0'$. Here we choose $L_x\times L_y=24\times 24$ for $\phi=1/3$, $1/4$, and $1/6$ and $L_x\times L_y=25\times 25$ for $\phi=1/5$. These values are not sensitive to the number and positions of well-separated holes and almost remain the same on larger lattices.}
\begin{center}
\begin{ruledtabular}
\begin{tabular}{cccc}
$\phi$ & $\tilde{t}_1/t_0$ & $\tilde{t}_2/t_0$ & $f$ \\ \hline
$1/3$ & $0.40$ & $0.30$ & $1.43$ \\ 
$1/4$ & $0.34$ & $0.19$ & $3.15$ \\
$1/5$ & $0.32$ & $0.14$ & $4.52$ \\ 
$1/6$ & $0.30$ & $0.10$ & $5.43$
\end{tabular}
\end{ruledtabular}
\end{center}
\end{table}

\section{Conclusion} 
\label{sec::conclusion}
In this work, we presented a microscopic study of FQH states with gapped boundaries in an extreme lattice limit. After gapping out boundaries, we found compelling signatures of the $\nu=k/2$ bosonic RR states with gapped boundaries residing on effective higher-genus surfaces. In particular, our numerical results strongly suggest the Hofstadter bilayers with holes are a promising platform to realize these novel FQH states. Excitingly, a scheme in optical lattices with spatially shaped laser beams generated by high resolution optics has been proposed to realize this setup in experiments~\cite{punchole}.

While there are many obvious variations of our model worth investigating in future works, the key message from this work is that even the most extreme lattice limit with high flux densities and single-site holes works remarkably well for realizing FQH states with gapped boundaries. This is particularly encouraging in the context of experimental realizations and possible future applications such as topological quantum computation since the energy scales, and thus the potential energy gap and associated critical temperatures, are maximal in this limit.

\begin{acknowledgments}
We thank G. M\"oller for related work. Z.L. thanks M. Hafezi and S. Simon for the helpful discussions. E.J.B. is supported by the Swedish Research Council (VR) and the Wallenberg Academy Fellows program of the Knut and Alice Wallenberg Foundation.  
\end{acknowledgments}

\appendix

\section{Project interactions to the higher-genus flat band}
\label{sec::proj}
For the model described as Eqs.~(\ref{eq::H0}) and (\ref{eq::H02}), the higher-genus flat band consists of the lowest $2\phi L_xL_y-M$ eigenstates of the single-particle Hamiltonian. Taking the two-body on-site interaction as an example, the interaction after band projection is
\begin{eqnarray}
&&H_{\mathrm{int}}^{\mathrm{proj}}=\sum_{m_1,m_2,m_3,m_4=1}^{2\phi L_xL_y-M}C_{m_1,m_2,m_3,m_4} a^\dagger_{m_1}a^\dagger_{m_2}a_{m_3}a_{m_4},\nonumber\\
&&C_{m_1,m_2,m_3,m_4}=\sum_{i=1}^{2L_xL_y}\psi^*_{m_1,i}\psi^*_{m_2,i}\psi_{m_3,i}\psi_{m_4,i},
\end{eqnarray} 
where $a^\dagger_{m}$ ($a_{m}$) creates (annihilates) a boson on the eigenstate $\psi_m$ of the single-particle Hamiltonian and $\psi_m=(\psi_{m,1},...,\psi_{m,2L_xL_y})$ is expressed in the lattice site basis. $\psi_{m,i}=0$ if the component $i$ corresponds to a removed lattice site. The projected three-body and four-body interactions used in this work can be derived similarly.

\section{Positions of holes} 
\label{sec::hpos}
Here we give the positions of minimal holes used to produce Figs.~\ref{fig::lau}, \ref{fig::nA}, \ref{fig::HS_lau}, and \ref{fig::HS_nA} (see Table~\ref{tab::holeposition}). We emphasize that different positions of holes give very similar results in both Abelian and non-Abelian cases as long as the holes are well separated.

\begin{table}
\caption{\label{tab::holeposition} The positions of minimal holes (i.e., the removed lattice sites) for various lattice sizes with $M=1$ and $M=2$, which are used to produce Figs.~\ref{fig::lau}, \ref{fig::nA}, \ref{fig::HS_lau}, and \ref{fig::HS_nA}.}
\begin{center}
\begin{ruledtabular}
\begin{tabular}{ccc}
$L_x\times L_y$ & $M=1$ & $M=2$ \\ \hline
$3\times 5$ & $(1,2)$ & \\
$4\times 4$ & $(1,1)$ & \\
$4\times 5$ & $(1,2)$ & \\
$5\times 4$ & $(2,1)$ & \\
$4\times 6$ & $(1,2)$ & $(1,1),(3,4)$ \\
$6\times 4$ & $(2,1)$ & $(1,1),(4,3)$ \\
$5\times 5$ & $(2,2)$ & \\
$4\times 7$ &  & $(1,1),(3,4)$ \\
$5\times 6$ &  & $(1,1),(3,4)$ \\
$6\times 5$ & $(2,2)$ & $(1,1),(4,3)$ \\
$8\times 4$ & $(3,1)$ & $(1,1),(5,3)$ \\
$5\times 7$ &  & $(1,1),(3,4)$ \\
$8\times 5$ & $(3,2)$ & $(1,1),(5,3)$ 
\end{tabular}
\end{ruledtabular}
\end{center}
\end{table}

\begin{figure*}[t]
\centerline{\includegraphics[width=\linewidth] {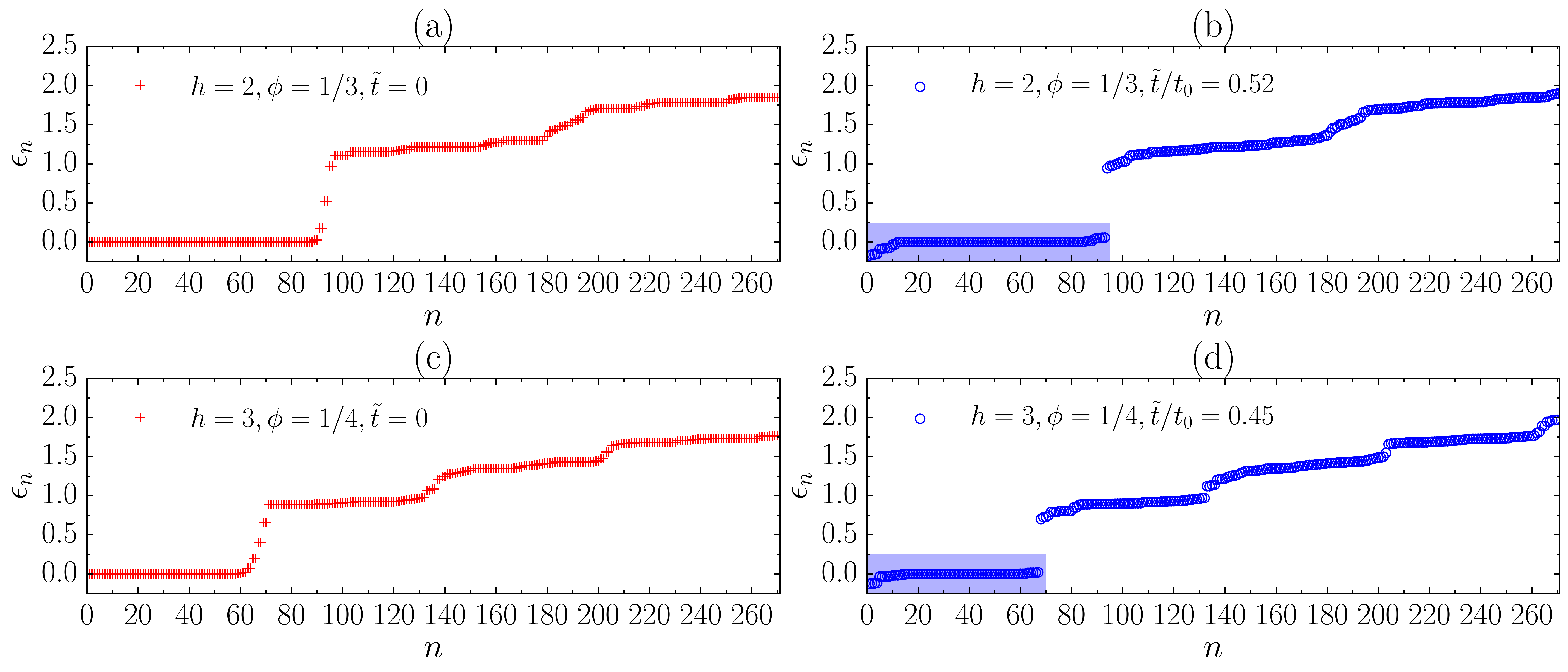}}
\caption{The spectra $\{\epsilon_n\}$ of Kapit-Mueller bilayers on an $L_x\times L_y=12\times 12$ lattice with one pair of larger holes. We have $h=2,\phi=1/3$ in (a) and (b) and $h=3,\phi=1/4$ in (c) and (d). The red pluses represent the spectra without interlayer tunneling. The blue circles represent the spectra with nonzero interlayer tunneling. For suitable tunneling strength $\tilde{t}$, we get a nearly flat lowest band (blue shading) in both cases. Here we use the optimal values of $\tilde{t}$ yielding the flattest lowest band.}
\label{fig::KM_SP_9h}
\end{figure*}

\section{Minimally entangled states} 
\label{sec::mes}
In Sec.~\ref{sec::fqh}, we study the MESs corresponding to different bipartitions of the $\nu=1/2$ Laughlin state on the effective $g=2$ surface emerging in the Kapit-Mueller bilayers with one pair of minimal holes. In order to search for the MESs in the ground-state manifold, we parametrize the superposition of the four degenerate ground states $|\Psi_{i=1,2,3,4}\rangle$ as
\begin{eqnarray} 
|\Psi(\{\theta_i\},\{\phi_i\})\rangle&=&\cos\theta_1|\Psi_1\rangle+\sin\theta_1\cos\theta_2 e^{i\phi_1}|\Psi_2\rangle\nonumber\\
&+&\sin\theta_1\sin\theta_2\cos\theta_3 e^{i\phi_2}|\Psi_3\rangle\nonumber\\
&+&\sin\theta_1\sin\theta_2\sin\theta_3 e^{i\phi_3}|\Psi_4\rangle,
\end{eqnarray} 
where $\theta_i\in[0,\pi/2]$ and $\phi_i\in[0,2\pi)$. For each type of bipartition, we start from a specific $|\Psi(\{\theta_i\},\{\phi_i\})\rangle$ to search for the point with the smallest entanglement entropy in the six-dimensional parameter space spanned by $\{\theta_i\}$ and $\{\phi_i\}$. When the minimization is approached, the superposition $|\Psi(\{\theta_i\},\{\phi_i\})\rangle$ is an MES (the von Neumann entropy and R\'enyi-$2$ entropy give similar results). In the presence of multiple MESs, each time we start the search from a $|\Psi(\{\theta_i\},\{\phi_i\})\rangle$ orthogonal to all MESs (with respect to the same bipartition) that have been found and examine whether we get a new MES when the minimization is approached. The global phase of each MES is fixed by requiring the first row and the first column of the overlap matrix $\mathcal{O}$ to be real. For each type of bipartition, we indeed obtain four MESs with similar R\'enyi-$2$ entropies $S_2$ (Table~\ref{tab::mes}), and the discrepancy between the four entropies is a finite-size effect and should disappear in the thermodynamic limit. 

The overlap matrix $\mathcal{O}$ between the MESs with respect to cuts I and II [Figs.~\ref{fig::lau_hc}(c) and \ref{fig::lau_hc}(d)] contains the information of the modular $\mathcal{S}$ matrix of the underlying topological order. For $N_b=6,L_x\times L_y=4\times 7,\phi=1/4$ and $N_b=6,L_x\times L_y=5\times 7,\phi=1/5$, we have shown that $\mathcal{O}$ is very close to 
\begin{eqnarray}
\mathcal{S}\otimes\mathcal{S}=
\frac{1}{2}
\left(
\begin{array}{cccc}
1&1&1&1\\
1&-1&1&-1\\
1&1&-1&-1\\
1&-1&-1&1
\end{array}\right),
\end{eqnarray}
where $\mathcal{S}=\frac{1}{\sqrt{2}}
\left(
\begin{array}{cc}
1&1\\
1&-1
\end{array}\right)
$
is the modular $\mathcal{S}$ matrix of the $\nu=1/2$ Laughlin state. This, together with the nonzero minimal entanglement entropy between two layers (Table~\ref{tab::mes}), identifies the ground state as the $\nu=1/2$ Laughlin state on the effective $g=2$ surface. For the smaller system size $N_b=6,L_x\times L_y=4\times 7,\phi=1/4$, the discrepancy of numerically extracted $\mathcal{O}$ from the theoretical value is larger compared to $N_b=6,L_x\times L_y=5\times 7,\phi=1/5$. This is due to the stronger finite-size effect on the smaller lattice, which is also reflected by the larger ground-state splitting [Fig.~\ref{fig::lau_hc}(a)].

It would be very interesting to study the MESs and extract the quasiparticle statistics also at non-Abelian filling factors. We have examined the situation for Kapit-Mueller bilayers with one pair of minimal holes at $\nu=1$. In this case, our calculations by real-space exact diagonalization are limited to only a few systems whose sizes are still small compared to those reachable by band projection, even if we have imposed a three-body hardcore condition. Unfortunately, we did not observe nice tenfold ground-state degeneracies in all systems that we checked by real-space exact diagonalization, which prevents us from studying the MESs and extracting the quasiparticle statistics for the MR state on a single $g=2$ surface. Considering that clear tenfold degeneracies exist in the spectra of the band-projected interaction [Fig.~\ref{fig::nA}(a)], we believe that this problem can be solved by seeking access to larger system sizes with other numerical techniques. Moreover, another technical difficulty here is the large ground-state degeneracy at non-Abelian filling factors, which leads to a complicated high-dimensional minimization when we search for MESs.

The stronger finite-size effects in Hofstadter bilayers than in Kapit-Mueller bilayers are also reflected in the calculations of MESs. Although we can already observe nice ground-state degeneracies by real-space exact diagonalization of small systems for the Kapit-Mueller bilayer model [Fig.~\ref{fig::lau_hc}(a)], such degeneracies are absent when we diagonalize the same system sizes in real space for the Hofstadter bilayer model, thus preventing us from studying the MESs and extracting the quasiparticle statistics. Therefore, we must seek numerical access to larger system sizes if we want to investigate the FQH physics with gapped boundaries in the Hofstadter bilayer model more deeply.

\section{Larger holes in Kapit-Mueller bilayers} 
\label{sec::lh}
So far we have focused on minimal holes, each of which contains only a single removed lattice site. Now we increase the hole size and study the single-particle physics of Kapit-Mueller bilayers with larger holes. 

We consider square holes each containing $h\times h$ removed lattice sites, such that the edge of each hole contains $4h+4$ sites. The single-particle Hamiltonian is still given by $H_0$ [Eq.~(\ref{eq::H0})], but now $\mathcal{R}$ contains $Mh^2$ removed sites in a single layer and $\mathcal{E}_m$ includes $4h+4$ edge sites of the $m$th hole in a single layer. For simplicity, we assume that the interlayer tunneling strength is $\tilde{t}$ for all of the $4h+4$ vertical tunneling terms between a pair of holes. The phases of the interlayer tunneling are again chosen to guarantee that each vertical plaquette between a pair of holes is pierced inwardly by effective flux $\phi$. 

In Fig.~\ref{fig::KM_SP_9h}, we show the band structure of Kapit-Mueller bilayers with $M=1$ pair of larger holes. We consider $h=2,\phi=1/3$ and $h=3,\phi=1/4$ on an $L_x\times L_y=12\times 12$ lattice. Without interlayer tunneling, one can see that there are more in-gap edge states than in the minimal hole case. However, a nearly flat lowest band can still be restored after interlayer tunneling is switched on to gap out the boundaries. The largest band flatness is $f=3.87$ and $f=4.64$ for $h=2,\phi=1/3$ and $h=3,\phi=1/4$, respectively. Compared with the numbers in Table~\ref{tab::t1t2}, the bands here are more dispersive. This is as expected because we use less fine tuning here. It would be interesting to study the many-body physics driven by interactions in the flat bands in the presence of larger holes.


\begin{thebibliography}{99}
\bibitem{tjk}
To adjust $t^\sigma_{jk}\equiv t^\sigma(x_j,y_j;x_k,y_k)$ to finite lattices under periodic boundary conditions, we replace it with $\sum_{s,t=-\infty}^{+\infty}t^\sigma(x_j+sL_x,y_j+tL_y;x_k,y_k)e^{-2\pi i s L_x y_j \phi_\sigma}$.
\end{thebibliography}
\end{document}